\documentclass[aps,prb,superscriptaddress,tightenlines,floatfix,notitlepage,twocolumn]{revtex4-1}
\usepackage{graphicx}
\usepackage[english]{babel}
\usepackage{amsmath}
\usepackage{amssymb}
\usepackage{tensor}

\newcommand{\be}{\begin{eqnarray}}
\newcommand{\ee}{\end{eqnarray}}
\newcommand{\p}{\partial}

\def\ket#1{|#1\rangle}
\def\bra#1{\langle #1 |}

\begin{document}

\title{Thermal Uhlmann Chern number from the Uhlmann connection for extracting topological properties of mixed states}

\author{Yan He}
\affiliation{College of Physical Science and Technology,
Sichuan University, Chengdu, Sichuan 610064, China}

\author{Hao Guo}
\affiliation{Department of Physics, Southeast University, Nanjing, Jiangsu 211189, China}

\author{Chih-Chun Chien}
\affiliation{School of Natural Sciences, University of California, Merced, CA 95343, USA.}
\email{cchien5@ucmerced.edu}

\begin{abstract}
The Berry phase is a geometric phase of a pure state when the system is adiabatically transported along a loop in its parameter space. The concept of geometric phase has been generalized to mixed states by the so called Uhlmann phase. However, the Uhlmann phase is constructed from the Uhlmann connection that possesses a well defined global section. This property implies that the Uhlmann connection is topologically trivial and as a consequence, the corresponding Chern character vanishes. We propose modified Chern character whose integral gives the thermal Uhlmann Chern number, which is related to the winding number of the mapping defined by the Hamiltonian. Therefore, the thermal Uhlmann Chern number reflects the topological properties of the underlying Hamiltonian of a mixed state. By including the temperature dependence in the volume integral, we also introduce the non-topological thermal Uhlmann Chern number which varies with temperature but is not quantized at finite temperatures. We illustrate the applications to a two-band model and a degenerate four-band model.
\end{abstract}

\maketitle

\section{introduction}
The discoveries of topological materials~\cite{Kane_TIRev,Zhang_TIRev,ShenTI,Bernevig_book,Chiu2016,Stanescu_book} have drawn attention to the topological properties of quantum matter. Most of the works have so far focused on the ground-state properties, such as the quantum anomalous Hall effect, topological insulators, etc.
The basis for understanding those topological properties is the concept of geometric phase introduced by Berry when studying the adiabatic evolutions in quantum mechanics~\cite{Berry}. The 1D Berry phase, also known as the Zak phase~\cite{Zak89}, has been measured in cold-atom systems~\cite{atala_direct_2013}. Since the geometric phase may be defined in classical systems as well~\cite{GPhase_book}, interesting topological properties have been studied in classical systems as well~\cite{Huber16}.

In a more mathematical language, the adiabatic condition provides a parallel condition between two different pure states in the Hilbert space~\cite{Stanescu_book,GPhase_book}. Starting from an initial state, one arrives at a parallel-transported state with a geometric phase after an adiabatic transportation. This geometric phase, called the Berry phase, can be computed as a line integral of a U(1) gauge connection defined for a specific eigenstate $|\psi(t) \rangle$ as $A=-i\bra{\psi(t)}\frac{\p}{\p t}\ket{\psi(t)}$. Here $t$ is a parameter in the Hilbert space. This notion can also be generalized to the non-abelian U(n) gauge connection for $n$ degenerate eigenstates $\psi_\mu(t)$ as $A_{\mu\nu}=-i\bra{\psi_\mu(t)}\frac{\p}{\p t}\ket{\psi_\nu(t)}$, where $\mu,\nu=1,2,\cdots,n$.

At finite temperatures, a quantum system is described by a mixed state~\cite{MQM}.
The notion of Berry connection leading to the Berry phase has been generalized to mixed states by Uhlmann's pioneer work~\cite{Uhlmann,Uhlmann1,Uhlmann2}. Ref.~\cite{Viyuela14} pioneered the Uhlmann's construction to define topological phases for mixed states, and Ref.~\cite{Viyuela18} demonstrated the measurement of Uhlmann's phase. For a mixed state, we consider a space of density matrices depending on a continuous parameter of the system. One can define a parallel condition in the space of density matrices as follows. First, a purification~\cite{IssacChuang_book,GPhase_book} of the density matrix $\rho$ is defined as $\rho=WW^\dagger$. Here the unitary matrix $W$ is called the Hilbert-Schmidt operator or the amplitude of the density matrix, which plays a similar role as $\ket{\psi}$ of the pure state.

As summarized in the Appendix, one can define a parallel condition for a parallel transport of the amplitude of a density matrix. The resulting infinitesimal form of the parallel transport gives rise to the Uhlmann connection. For a $n$ by $n$ density matrix $\rho$ with the spectral decomposition $\rho=\sum_{i=1}^{n}p_i\ket{i}\bra{i}$, the Uhlmann connection is given by
\be
A_U=\sum_{i,j}\ket{i}\bra{i}\frac{[d\sqrt{\rho},\,\sqrt{\rho}]}{p_i+p_j}\ket{j}\bra{j}.
\label{AU}
\ee
Here $d$ denotes the derivative with respect to the continuous parameter and $[A,B]$ denotes the commutator of $A$ and $B$. The derivation of Eq.~(\ref{AU}) is summarized in the Appendix. One can verify that Eq.~(\ref{AU}) defines the connection of a $U(n)$ bundle over the parameter space of the Hamiltonian.
With this connection, the $U(n)$ holonomy of a curve $\gamma$ reads
\be
U_\gamma=\mathcal{P}\exp\Big(\int_\gamma A_U\Big).
\ee
Here $\mathcal{P}$ stands for the path ordering along $\gamma$. Thus, if we consider the holonomy of a closed loop $c$, one can define the so-called Uhlmann phase $\Phi_U$ via
\be
\exp(i\Phi_U)=\mbox{tr}(\rho(0) U_c). \label{PhiU}
\ee
Here $\rho(0)$ is the density matrix of the initial point of the path and $\mbox{tr}$ denotes the trace. 

Based on the Uhlmann connection, it was proposed in Ref.~\cite{Huang14} that by studying the eigenvalues of the holonomy matrix of homological non-trivial loops of a 2D model, one can reveal a topological transition at finite $T$. It is also suggested in Ref.~\cite{Diehl} that by constraining the Hilbert-Schmidt operator space, one can achieve a topologically non-trivial Uhlmann connection to describe the topology of mixed states. Ref.~\cite{Mera17} studied the Uhlmann connection in fermionic systems undergoing phase transitions. We also briefly mention some other approaches that do not use the Uhlmann connection: Ref.~\cite{Diehl18} proposed the so-called ensemble geometric phase based on the generalization of the polarization of a lattice model to finite $T$. Ref.~\cite{Bardyn18} considered the homotopic extension of a topological observable of a pure state to a mixed state. Ref.~\cite{Sjoqvist00} obtained a geometric phase of mixed state based on a generalization of Pancharatnam's connection.

In the following, we will briefly review the challenges of extracting topological properties from the Uhlmann connection by using available approaches. To offer an alternative solution to the challenges, we propose another approach where the Uhlmann connection is adopted but the definition of the Chern character is generalized. We will introduce the thermal Uhlmann Chern numbers characterizing the topological properties arising from the mapping between the parameter space and target space of the Hamiltonian. Two generic systems, one in two-dimension and one in four-dimension, are analyzed to verify the thermal Uhlmann Chern number indeed captures the underlying topological properties.

\section{Challenge of extracting topological properties of mixed states}
Although the Uhlmann phase may be viewed as an analogue of the Berry phase for mixed states, it actually has some known drawbacks. In its definition, $\rho$ is required to be non-singular or without any zero eigenvalues. This implies that one cannot directly apply Eq.~(\ref{AU}) to pure states. Nevertheless, one can still verify that, with extra care, $A_U$ can approach the corresponding Berry connection as $T\to0$ \cite{Viyuela14,Viyuela15}.
A more severe problem of the Uhlmann connection is that the density matrix bundle is topologically trivial. Since $\sqrt{\rho}$ is non-singular and $p_i$ is always nonzero (when pure states are excluded), it follows that $A_U$ is always non-singular. This means that $A_U$ is a global section of the $U(n)$ bundle, which then implies that this bundle is topologically trivial. As a consequence, all characteristic classes, such as the Chern class and Chern character, are all zero~\cite{Viyuela15}. This implies that even though the Uhlmann connection approaches the Berry connection as $T\rightarrow 0$, the underlying bundle structures can still be different. Therefore, one cannot find any nontrivial topological information in the strict mathematical sense from the Ulhmann connection.

Nevertheless, some alternative methods for extracting topological information of mixed states have been proposed. For example, Ref.~\cite{Viyuela2} proposed the Uhlmann number, which was argued to approach the Chern number as $T\to 0$. The Uhlmann number is obtained by slicing the Brillouin zone (BZ) into a bunch of circles. One first calculates the Uhlmann phase $\Phi_U(k_x)$ along each $k_y$ circle in the BZ with a fixed value of $k_x$. Then, the Uhlmann number is given by integrating over $k_x$:
\be
C_U=\frac{1}{2\pi}\int_{S_1}\frac{\p \Phi_U(k_x)}{\p k_x}d k_x.
\label{CU}
\ee
One can also compute the $k_x$ loop at a fixed value of $k_y$ to find $\Phi_U(k_y)$, and then define $\tilde{C}_U=\frac{1}{2\pi}\int_{S_1}\frac{\p \Phi_U(k_y)}{\p k_y}d k_y$.
Importantly, it has been shown in Section IV of Ref.~\cite{Diehl} that those two definitions in general do not agree with each other. Therefore, the Uhlmann number may not serve as an intrinsic quantity of the system. The reason is that although the holonomy has the $U(n)$ group structure, the $\phi_U$ defined in Eq.~(\ref{PhiU}) does not have a $U(1)$ additive group structure. This renders the inconsistent definitions of $C_U$.

An alternative approach~\cite{Diehl,Bardyn18} has also been proposed to find a topologically nontrivial bundle by constraining the Hilbert-Schmidt operator space. Instead of taking the full-rank density matrix of the whole system, one considers its projection to a subspace of the whole Hilbert space. The resulting singular density matrix is then expanded by a set of nonorthogonal states. Then, by following a similar derivation of the Uhlmann connection, one can find a different but topologically nontrivial connection. The topological invariants obtained by this method will retain the same values for all finite temperatures and only vanish as $T\to\infty$. As a consequence, there is no finite-temperature transition of the topological invariants.

Here, we take the original Uhlmann connection $A_U$ but in order to extract topological information from it, we propose to modify the formula of the Chern character and derive the thermal Uhlmann Chern number based on $A_U$. One advantage of the approach is that the thermal Uhlmann Chern number is obtained from a volume integral instead of an integral of accumulated Uhlmann phase as shown in Eq.~(\ref{CU}). Thus, the resulting thermal Uhlmann Chern number does not depend on how one decomposes the BZ and reflects intrinsic properties of the system. Importantly, the thermal Uhlmann Chern number is related to the winding number of the mapping defined by the Hamiltonian. As a consequence, the thermal Uhlmann Chern number is manifestly quantized and has the same value as the Chern number computed from the Berry connection at zero temperature.

\section{Thermal Uhlmann Chern number of a two-band model}
We consider a generic two-band Hamiltonian $H=R_i\sigma_i$. Here $\sigma_i$ with $i=1,2,3$ denote the Pauli matrices and $R_i$ are real-number functions with their arguments from a 2D parameter space (say, spanned by $k_x$ and $k_y$). Here the repeated indices imply a summation (the Einstein convention) and we assume the Hamiltonian is dimensionless. The two eigenstates $\ket{u_{1,2}}$ are given by
\be
\ket{u_{1,2}}=\frac{1}{\sqrt{R(R\pm R_3)}}\left(\begin{array}{c}
                                         R\pm R_3 \\
                                         R_1+i R_2
                                       \end{array}\right).
\ee
Here the eigenvalues are $\pm R$ and  $R=\sqrt{R_i^2}$. The corresponding Boltzmann weights are given by
\be
p_{1,2}=\frac{e^{\pm R/T}}{Z}
\ee
with the partition function $Z=2\cosh(R/T)$.
For later convenience, we also introduce the projection operators as 
\be
P_{1,2}=\ket{u_{1,2}}\bra{u_{1,2}}=\frac12(1\pm\hat{R}_i\sigma_i).
\ee
They lead to $P_1\ket{u_1}=\ket{u_1}$, $P_1\ket{u_2}=0$, etc. Here $\hat{R_i}=R_i/R$. Then, the density matrix can be expressed as
\be
\rho=p_i P_i=\frac12\Big(1+\tanh(\frac{R}{T})\hat{R}_i\sigma_i\Big).\label{rho}
\ee

The Uhlmann connection $A_U$ and its curvature can be expressed by the differential forms. Explicitly,
\be
A_U&=&f(R)\Big(\ket{u_1}\bra{u_1}(d\ket{u_2})\bra{u_2}+\ket{u_2}\bra{u_2}(d\ket{u_1})\bra{u_1}\Big)\nonumber\\
&=&f(R)(P_1dP_2+P_2dP_1)=-\frac12f(R)\hat{R}_id\hat{R}_j\sigma_i\sigma_j\nonumber\\
&=&-\frac{i}2f(R)\epsilon_{ijk}\hat{R}_id\hat{R}_j\sigma_k .
\ee
Here $f(R)=1-\frac{1}{\cosh(R/T)}$. In the derivation we have used $\sigma_i\sigma_j=\delta_{ij}+i\epsilon_{ijk}\sigma_k$ with $\epsilon_{ijk}$ being the Levi-Civita symbol, and $\hat{R}_id\hat{R}_i=0$.
The Uhlmann curvature is defined as $F_U=dA_U+A_U\wedge A_U$. These two terms are found to be
\be
&&dA_U=-\frac{i}2f'(R)dR\wedge\epsilon_{ijk}\hat{R}_id\hat{R}_j\sigma_k\nonumber\\
&&\qquad-\frac{i}2f(R)\epsilon_{ijk}d\hat{R}_i\wedge d\hat{R}_j\sigma_k\label{dA}, \\
&&A_U\wedge A_U=-\frac{i}{4}f^2(R)\epsilon_{ijk}d\hat{R}_i\wedge d\hat{R}_j\sigma_k .
\ee
Here $f'(R)=\partial f/\partial R$.

Both $A_U$ and $F_U$ are traceless by definition. As a consequence, the Chern character of the Uhlmannn connection is trivially zero, and this leads to a vanishing Chern number. Explicitly,
\be
Ch_1=\frac{i}{2\pi}\int \mbox{tr}F_U=0.
\ee
In order to obtain nontrivial results associated with the $T=0$ topological properties, we insert the density matrix into the definition of the Chern character and find the following identity:
\be
&&\mbox{tr}(\rho F_U)=-\frac{i}{4}\epsilon_{abc}\hat{R}_a d\hat{R}_b\wedge d\hat{R}_c\nonumber\\
&&\quad\times\tanh(\frac RT)\Big[2f(R)-f^2(R)\Big]\label{rF}.
\ee
Note that the first term of $dA_U$ in Eq.~(\ref{dA}) does not contribute because $\epsilon_{ijk}\hat{R}_id\hat{R}_j\hat{R}_k=0$.

At low $T$, $\tanh(R/T)\to1$ and $f(R)\to1$, so $\mbox{tr}(\rho F_U)$ approaches the volume form of the target space $S^2$ defined by $\hat{R}_i$. Because of Eq.~(\ref{rF}), we introduce the following thermal Uhlmann Chern number.
\begin{eqnarray}
\widetilde{Ch}_1&=&\frac{i}{2\pi}\int\lambda_1(R,T)\mbox{tr}(\rho F_U) \nonumber \\
&=&\frac{1}{8\pi}\int\epsilon_{abc}\hat{R}_a d\hat{R}_b\wedge d\hat{R}_c
\label{Ch1}
\end{eqnarray}
with $1/\lambda_1(R,T)=\tanh(\frac RT)\Big[2f(R)-f^2(R)\Big]$. The thermal Uhlmann Chern number (\ref{Ch1}) is expressed as a 2D volume integral in the parameter space. This guarantees the results are independent of the order of integrations of different parameters, similar to the topological invariants of mixed states defined in Refs.~\cite{Diehl,Bardyn18}. Moreover, one can see that the thermal Uhlmann Chern number is related to the winding number of the mapping from the parameter space to the target space.

To visualize the implications of the thermal Uhlmann Chern number, we consider a specific 2D tight-binding model usually known as the Qi-Wu-Zhang model~\cite{Qi1}.  In this model, the three components of $R_i$ are taken to be $R_1=\sin k_x$, $R_2=\sin k_y$ and $R_3=m+\cos k_x+\cos k_y$. After some algebra, the thermal Uhlmann Chern number at finite $T$ is given by
\be
\widetilde{Ch}_1=\left\{
     \begin{array}{ll}
       1, & 0<m<2, \\
       -1, & -2<m<0, \\
       0, & |m|>2 ,
     \end{array}
   \right.
\ee
which is identical to the Chern number obtained from the Berry connection at $T=0$. Thus, their values can be inferred if the $T=0$ Chern number is known. Nevertheless, recent developments on mixed-state observables~\cite{Sjoqvist00,Bardyn18,Diehl18} may lead to direct measurements of the thermal Uhlmann Chern number.

While previous studies using the Uhlmann number~\cite{Viyuela14} or topological index associated with the Uhlmann connection~\cite{Huang14} predict possible topological transitions at finite temperatures, here the thermal Uhlmann Chern number takes the same value for any finite temperature. We caution that when $T\to\infty$, the density matrix becomes the identity matrix and one can no longer differentiate the weights of the underlying states. As a consequence, $A_U=0$ and $\widetilde{Ch}_1=0$.

\begin{figure}
\includegraphics[width=0.48\textwidth]{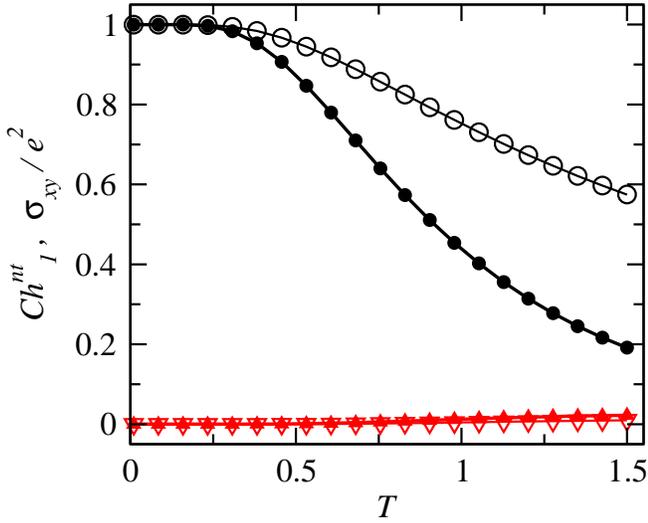}
\caption{The non-topological (NT) thermal Uhlmann Chern number (solid symbols) from Eq.~\eqref{eq:ntch1} and $\sigma_{xy}/e^2$ (hollow symbols) from Eq.~\eqref{eq:tcond} of the two-band model as functions of temperature. The red and black symbols correspond to $m=1.3$ and $m=2.3$, respectively.}
\label{TCh1}
\end{figure}

In the derivation of the thermal Uhlmann Chern number, the thermal factor can be rewritten as $1/\lambda_1(R,T)=\tanh^3(\frac RT)$. This suggests that the right hand side of Eq. (\ref{rF}) can be expressed as a volume form with the unit vector $\hat{R}_a$ replaced by $\hat{R}_a\tanh(\frac RT)$. In the presence of the thermal factor, however, the integral of the volume form is not quantized. Therefore, we introduce a non-topological (NT) thermal Uhlmann Chern number as~\cite{Referee1}
\be\label{eq:ntch1}
\widetilde{Ch}_1^{nt}&=&\frac{i}{2\pi}\int\mbox{tr}(\rho F_U) \nonumber \\
&=&\frac{1}{8\pi}\int\epsilon_{abc}\hat{R}_a d\hat{R}_b\wedge d\hat{R}_c\tanh^3(\frac RT).
\ee
Note that $\hat{R}_a\tanh(R/T)$  also appears in the density matrix, Eq. (\ref{rho}).

In the two-band model, the NT thermal Uhlmann Chern number has a qualitatively similar dependence on temperature as the transverse conductivity~\cite{ShenTI}. The latter at finite $T$ is given by
\be
\sigma_{xy}=\frac{e^2}{2\pi}\sum_{n=1}^2\int n_F(E_n)F^n.
\ee
Here $n=1,2$ label the two bands, $E_n$ denotes the energy of the $n$-th band, $n_F(E)$ is the Fermi distribution. $F^n=dA^n$ and $A^n=-i\bra{u_n}(d\ket{u_n})$ are the Berry curvature and Berry connection of the $n$-th band. Making use of the following identity for the two band model,
\be
F^1=-F^2=\frac{1}{4\pi}\epsilon_{abc}\hat{R}_a d\hat{R}_b\wedge d\hat{R}_c,
\ee
and the energy $E_1=-E_2=-R$, we find that
\be\label{eq:tcond}
\sigma_{xy}=\frac{e^2}{8\pi}\int\epsilon_{abc}\hat{R}_a d\hat{R}_b\wedge d\hat{R}_c\tanh(\frac RT).
\ee
Since the tanh function goes from $1$ to $0$ as $T$ increases from $0$ to infinity, $\sigma_{xy}$ and the NT thermal Uhlmann Chern number have similar $T$ dependence. Therefore, the transverse conductivity at finite $T$ may act as a proxy for the NT thermal Uhlmann Chern number or vice versa.  Ref.~\cite{Rivas13} also presents a discussion of the density-matrix Chern number and transverse conductivity using a different approach.

For the Qi-Wu-Zhang model, we find the explicit expression of the non-topological (NT) thermal Uhlmann Chern number as
\be\label{eq:NTCh_2band}
& &\widetilde{Ch}_1^{nt}=-\frac1{4\pi}\int dk_x dk_y\nonumber\\
&\times&\frac{\cos k_x+\cos k_y+m\cos k_x\cos k_y}
{\Big[\sin^2 k_x+\sin^2 k_y+(m+\cos k_x+\cos k_y)^2\Big]^{3/2}}.
\ee
The numerical values of the non-topological thermal Uhlmann Chern number as a function of $T$ are shown in Figure \ref{TCh1}. The red triangles (black dots) represents the $m=1.3$ ($m=2.3$) case, which is topologically non-trivial (topologically trivial) at $T=0$ with the Chern number $Ch=1$ ($Ch=0$). Also shown in Figure \ref{TCh1} are $\sigma_{xy}/e^2$ (red upside-down triangles and black hollow dots) for the corresponding cases.

\section{Second thermal Uhlmann Chern number of a four-band model}
Next, we consider a more complicated four-band model. The Hamiltonian is
$H=\sum_{i=1}^5 R_i\Gamma^i$. Here the gamma matrices are $\Gamma_i=\sigma_1\otimes\sigma_i$ for $i=1,2,3$,  $\Gamma_4=\sigma_2\otimes I$, $\Gamma_5=\sigma_3\otimes I$, and $I$ is the identity matrix with suitable dimension. The gamma matrices satisfy $\{\Gamma_i,\Gamma_j\}=2\delta_{ij}$. The arguments of $R_i$ are from a 4D parameter space (say, $k_x$, $k_y$, $k_z$, and $k_u$).
The four eigenstates are given by
\be
&&\ket{u_{a,c}}=\frac{1}{\sqrt{2R(R\mp R_5)}}\Big(-R_3+i R_4,\nonumber\\
&&\qquad-R_1-iR_2,\,R_5\mp R,\,0\Big)^t, \\
&&\ket{u_{b,d}}=\frac{1}{\sqrt{2R(R\mp R_5)}}\Big(-R_1+i R_2,\nonumber\\
&&\qquad R_3+iR_4,\,0,\,R_5\mp R\Big)^t.
\ee
Here $R=\sqrt{R_i^2}$ and the superscript $t$ denotes the transpose. There are two doubly degenerate energy levels $H\ket{u_{a,b}}=R\ket{u_{a,b}}$ and $H\ket{u_{c,d}}=-R\ket{u_{c,d}}$. The Boltzmann weights of the two degenerate subspaces are
\be
p_{1,2}=\frac{2e^{\pm R/T}}{Z}
\ee
with the partition function $Z=4\cosh(R/T)$.
The projection operators of the subspaces and the density matrix are given by
\be
&&P_1=\ket{u_a}\bra{u_a}+\ket{u_b}\bra{u_b}=\frac12(1+\hat{R}_i\Gamma_i), \\
&&P_2=\ket{u_c}\bra{u_c}+\ket{u_d}\bra{u_d}=\frac12(1-\hat{R}_i\Gamma_i), \\
&&\rho=p_i P_i=\frac12\Big(1+\tanh(\frac{R}{T})\hat{R}_i\Gamma_i\Big).
\ee
Here $\hat{R_i}=R_i/R$.

The Uhlmann phase $A_U$ is found to be
\be
A_U&=&f(R)(P_1dP_2+P_2dP_1)=-\frac12f(R)\hat{R}_a d\hat{R}_b\Gamma_a\Gamma_b\nonumber\\
&=&\frac{i}2f(R)\Big(\hat{R}_a d\hat{R}_b-\hat{R}_b. d\hat{R}_a\Big)\Gamma_{ab}.
\ee
Here $f(R)=1-\frac{1}{\cosh(R/T)}$ and $\Gamma_{ab}=i[\Gamma_a, \Gamma_b]/2$ satisfying $\Gamma_{ab}^2=I$.
We have used the identity $\hat{R}_a d\hat{R}_a=0$.
Again, the Uhlmannn curvature is defined as $F_U=dA_U+A_U\wedge A_U$. These two terms are given by
\be
&&d A_U=-f'(R)\hat{R}_{[a} d\hat{R}_{b]}\Gamma_{ab}\wedge dR\nonumber\\
&&\quad +if(R)d\hat{R}_a\wedge d\hat{R}_b\Gamma_{ab}, \\
&&A_U\wedge A_U=\frac{i}2f^2(R)d\hat{R}_a\wedge d\hat{R}_b\Gamma_{ab}.
\ee
Here we define $\hat{R}_{[a} d\hat{R}_{b]}=\hat{R}_a d\hat{R}_b-\hat{R}_b d\hat{R}_a$.
The second Chern character is given by
\be
&&F_U\wedge F_U=-\frac14g^2(R)d\hat{R}_a d\hat{R}_b d\hat{R}_c d\hat{R}_d\Gamma_{ab}\Gamma_{cd}\nonumber\\
&&\quad+\Big[f'(R)\Big]^2\hat{R}_{[a} d\hat{R}_{b]}\Gamma_{ab}\hat{R}_{[c} d\hat{R}_{d]}\Gamma_{cd}\nonumber\\
&&\quad-\frac{i}2g(R)f'(R)\Big[\hat{R}_{[c} d\hat{R}_{d]}\Gamma_{cd} d\hat{R}_a d\hat{R}_b\Gamma_{ab}\nonumber\\
&&\quad+d\hat{R}_a d\hat{R}_b\Gamma_{ab}\hat{R}_{[c} d\hat{R}_{d]}\Gamma_{cd}\Big].
\label{FF}
\ee
Here $g(R)=2f(R)-f^2(R)$ and we omit the ``$\wedge$'' symbol.
It can be shown that tr$(F_U\wedge F_U)=0$. Thus, the second Chern number of the Uhlmann connection is trivially zero:
\be
Ch_2=\frac{1}{8\pi^2}\int \mbox{tr}(F_U\wedge F_U)=0.
\ee

Similar to the discussion of the thermal Uhlmann Chern number in the two-band case, we insert the density matrix into the second Chern character to obtain the following identity
\be
&&\mbox{tr}(\rho F_U\wedge F_U)=-\frac{3}{4!}\int\epsilon_{abcde}\hat{R}_a d\hat{R}_b d\hat{R}_c d\hat{R}_d d\hat{R}_e\nonumber\\
&&\quad\times\tanh(\frac RT)g^2(R).\label{rFF}
\ee
Note that the second and third terms of Eq.~(\ref{FF}) do not contribute.
At low $T$, $\tanh(R/T)\to1$ and $g(R)\to1$, we find that $\mbox{tr}(\rho F_U\wedge F_U)$ approaches the volume form of the target space $S^4$ defined by $\hat{R}_i$. Thus, we introduce the following second thermal Uhlmann Chern number, which is also equal to the winding number from the parameter space to the target space.
\begin{eqnarray}
\widetilde{Ch}_2&=&\frac{1}{8\pi^2}\int \lambda_2(R,T)\mbox{tr}(\rho F_U\wedge F_U) \nonumber \\
&=&\frac{3}{8\pi^2\cdot4!}\int\epsilon_{abcde}\hat{R}_a d\hat{R}_b d\hat{R}_c d\hat{R}_d d\hat{R}_e
\label{Ch2}
\end{eqnarray}
with $1/\lambda_2(R,T)=-\tanh(\frac RT)g^2(R)$. The second thermal Uhlmann Chern number (\ref{Ch2}) is expressed as a 4D volume integral in the parameter space. This guarantees the results are independent of the order of integrations of different parameters. The second Chern number also remains constant for any finite temperature, but it vanishes when $T\rightarrow\infty$ because the density matrix becomes the identity matrix. Again, this is similar to the topological invariants of mixed states defined in Refs.~\cite{Diehl,Bardyn18}.

Similar to the 2D case considered in the previous section, here we consider a specific 4D tight-binding model proposed to study the 4D quantum Hall effects~\cite{Qi2}.  In this model, the five components of $R_i$ are taken to be $R_1=\sin k_x$, $R_2=\sin k_y$, $R_3=\sin k_z$, $R_4=\sin k_u$, and $R_5=m+\cos k_x+\cos k_y+\cos k_z+\cos k_u$. The second thermal Uhlmann Chern number at finite $T$ is given by
\be
\widetilde{Ch}_2=\left\{
     \begin{array}{ll}
       3, & 0<m<2; \\
       -3, & -2<m<0; \\
       -1, & 2<m<4; \\
       1, & -4<m<-2; \\
       0, & |m|>4.
     \end{array}
   \right.
\ee
The values are the same as the second Chern number from the Berry connection at $T=0$, and the methods discussed in Refs.~\cite{Sjoqvist00,Bardyn18,Diehl18} may lead to direct measurements of the second thermal Uhlmann Chern number.
Moreover, the second thermal Uhlmann Chern number has a similar temperature dependence as the previously discussed first thermal Uhlmann Chern number.

The result of this section shows that the construction of the thermal Uhlmann Chern number can also be applied to systems with band degeneracy. Although a 4D model may not be realized experimentally in a straightforward way, through the method of dimensional reduction~\cite{Qi2}, it can be used to compute the $Z_2$ index of 3D topological insulators with time reversal symmetry.

\begin{figure}
\includegraphics[width=0.48\textwidth]{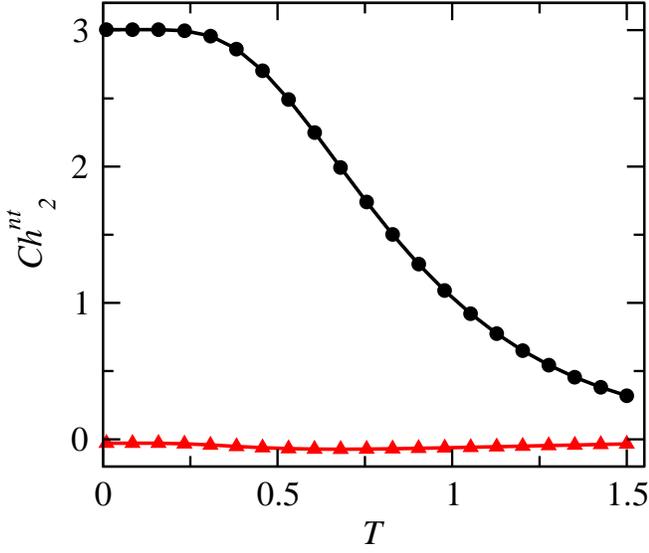}
\caption{The second NT thermal Uhlmann Chern number of the four-band model, Eq.~\eqref{eq:NTCN_4band}, as a function of temperature. The red triangles and black dots correspond to $m=1.3$ and $m=4.3$, respectively.}
\label{TCh2}
\end{figure}

The second thermal Uhlmann Chern number at any finite temperature again reflects the second Chern number at $T=0$. Similar to the derivation of the NT thermal Uhlmann Chern number of the two-band model, the right hand side of Eq. (\ref{rFF}) can be expressed as a volume form with the unit vector $\hat{R}_a$ replaced by $\hat{R}_a\tanh(\frac RT)$. The integral of this volume form, similar to the integral of the two-band model, is not quantized. We thus introduce the second NT Chern number as
\be
\widetilde{Ch}_2^{nt}&=&\frac{1}{8\pi^2}\int\mbox{tr}(\rho F_U\wedge F_U) \nonumber \\
&=&-\frac{3}{8\pi^2\cdot4!}\int\epsilon_{abcde}\hat{R}_a d\hat{R}_b d\hat{R}_c d\hat{R}_d d\hat{R}_e
\tanh^5(\frac RT).\nonumber \\
& &
\ee
For the 4D tight-binding model, the explicit expression of the second NT thermal Uhlmann Chern number is
\be\label{eq:NTCN_4band}
\widetilde{Ch}_1^{nt}&=&-\frac3{8\pi^2}\int \frac{dk_x dk_y dk_z dk_u}{R^5}
\Big(\cos k_x\cos k_y\cos k_z\nonumber\\
&+&\cos k_x\cos k_z\cos k_u+\cos k_y\cos k_z\cos k_u\nonumber\\
&+&m\cos k_x\cos k_y\cos k_z\cos k_u\Big).
\ee
The values of the second NT Chern number as a function of $T$ are shown in Figure \ref{TCh2}. The red triangles (black dots) show the $m=1.3$ ($m=4.3$) case, which is topologically non-trivial (topologically trivial) at $T=0$ with the Chern number $Ch=3$ ($Ch=0$).

\section{conclusion}
The challenges of generalizing the topological invariants to mixed states have been summarized. The Uhlmann connection leads to a trivial bundle structure, so its associated topological invariants vanish. Other proposals have their own issues, too. To extract topological information associated with the Uhlmann connection, the thermal Uhlmann Chern numbers are constructed by inserting the density matrix into the expressions of the Chern characters. The thermal Uhlmann Chern numbers are consistent with the zero-temperature Chern numbers and survive at finite temperatures. The correspondence between the thermal Uhlmann Chern number and the winding number clarifies their topological origin. In addition, the non-topological thermal Uhlmann Chern numbers can be defined by including the thermal factors into the volume integrals, albeit they are not quantized at finite temperatures. For the two-band model, the non-topological thermal Uhlmann Chern number is qualitatively similar to the transverse conductivity. Our proposal thus offers an alternative route to the characterization of topological properties of mixed states.

\textbf{Acknowledgment:} Y.H. is supported by NSFC under Grant No. 11404228. We thank the two anonymous Referees for many valuable suggestions. 

\appendix

\section{The calculation of Uhlmann connection $A_U$}
Here is a brief summary of the Uhlmann connection following Ref.~\cite{Hubner}. The concept of the Uhlmann connection is based on the amplitude decomposition of the density matrix:
\be
\rho=WW^{\dagger},\qquad W=\sqrt{\rho}U.
\ee
The parallel condition of a pair of different states is
\be
W_1^{\dagger}W_2=W_2^{\dagger}W_1=C>0
\ee
Here $C>0$ means $C$ is a positive definite matrix.
If we take $W_1=\sqrt{\rho_1}$, then we find
\be
&&C^2=W_1^{\dagger}W_2W_2^{\dagger}W_1=\sqrt{\rho_1}\rho_2\sqrt{\rho_1},\\
&&W_2=\sqrt{\rho_1^{-1}}\sqrt{\sqrt{\rho_1}\rho_2\sqrt{\rho_1}}.
\ee
The corresponding unitary transformations are
\be
U_1=1,\quad U_2=\sqrt{\rho_2^{-1}}\sqrt{\rho_1^{-1}}\sqrt{\sqrt{\rho_1}\rho_2\sqrt{\rho_1}}.
\ee
By choosing $\rho_1=\rho$ and $\rho_2=\rho+td\rho$, we have
\be
(U+tdU)U^{\dagger}=\sqrt{(\rho+td\rho)^{-1}}\sqrt{\rho^{-1}}
\sqrt{\sqrt{\rho}(\rho+td\rho)\sqrt{\rho}}. \nonumber
\ee
Then, the Uhlmann connection $A_U= dU U^{\dagger}$ becomes
\be
&&A_U=\frac{d}{dt}\sqrt{(\rho+td\rho)^{-1}}~\Big|_{t=0}\sqrt{\rho}\nonumber\\
&&\quad+\rho^{-1}\frac{d}{dt}\sqrt{\sqrt{\rho}(\rho+td\rho)\sqrt{\rho}}~\Big|_{t=0}.
\ee
Note that $A_U$ defined in this way is anti-Hermitian.

In order to find an explicit expression of $A_U$, we start with the Boltzmann weight and density matrix:
\be
p_i=\frac{e^{-E_i/T}}{\sum_i e^{-E_i/T}},\quad \rho=\sum_ip_i\ket{i}\bra{i}.
\ee
Let $A=\sqrt{(\rho+td\rho)}$ and $B=\sqrt{\sqrt{\rho}(\rho+td\rho)\sqrt{\rho}}$, then we find the following relations.
\be
&&\bra{i}\frac{d}{dt}A^2\ket{j}\Big|_{t=0}=(\sqrt{p_i}+\sqrt{p_j})\bra{i}\frac{d}{dt}A\ket{j}=\bra{i}d\rho\ket{j}, \nonumber\\
&&\bra{i}\frac{d}{dt}B^2\ket{j}\Big|_{t=0}=(p_i+p_j)\bra{i}\frac{d}{dt}B\ket{j}=\sqrt{p_ip_j}\bra{i}d\rho\ket{j}, \nonumber\\
&&\bra{i}\frac{d}{dt}A^{-1}\ket{j}=-\bra{i}A^{-1}\frac{dA}{dt}A^{-1}\ket{j}=-\frac{1}{\sqrt{p_ip_j}}\bra{i}\frac{dA}{dt}\ket{j}. \nonumber
\ee
From those identities, we find
\be
\bra{i}A_U\ket{j}&=&\sqrt{p_j}\bra{i}\frac{d}{dt}A^{-1}\ket{j}+p_i^{-1}\bra{i}\frac{d}{dt}B\ket{j}\nonumber\\
&=&-\frac{1}{p_i+p_j}\frac{\sqrt{p_i}-\sqrt{p_j}}{\sqrt{p_i}+\sqrt{p_j}}\bra{i}d\rho\ket{j}\nonumber\\
&=&\frac{\bra{i}[d\sqrt{\rho},\,\sqrt{\rho}]\ket{j}}{p_i+p_j}\label{AU1}.
\ee
Note that for the two band model discussed in the main text, $p_1+p_2=1$, then Eq.~(\ref{AU1}) can be simplified to  $A_U=[d\sqrt{\rho},\,\sqrt{\rho}]$.
In the more generic case, we can also rewrite $A_U$ as
\be
\bra{i}A_U\ket{j}
&=&\frac{1}{p_i+p_j}\frac{\sqrt{p_i}-\sqrt{p_j}}{\sqrt{p_i}+\sqrt{p_j}}(p_i-p_j)\bra{i}(d\ket{j})\nonumber\\
&=&\frac{(\sqrt{p_i}-\sqrt{p_j})^2}{p_i+p_j}\bra{i}(d\ket{j})
\ee
Here $d$ is the exterior derivative.


\begin{thebibliography}{31}
\bibitem [{Ref()}]{Referee1}%
	\BibitemOpen
	\href@noop {} {}\bibinfo {note} {The derivation was suggested by an anonymous
		referee.}\BibitemShut {Stop}%
\end{thebibliography}
%

\end{document}